\documentclass{PoS}
\pdfoutput=1
\usepackage{amsmath,amssymb}
\usepackage{graphicx}

\usepackage[sort&compress,numbers,merge]{natbib}
\usepackage{natbib,natbibspacing}
\usepackage{setspace}

\newcommand{\Fig}[1]{\mbox{Fig.\,\ref{#1}}}
\newcommand{\Tab}[1]{\mbox{Table\,\ref{#1}}}

\title{Nucleon generalized form factors and sigma term from lattice QCD near the
physical quark mass}

\ShortTitle{Nucleon generalized form factors and $\sigma_{\pi N}$ term near the
physical quark mass}

\author{G.S.~Bali, S.~Collins, B.~Gl{\"a\ss{}}le,
M.~G\"ockeler, J.~Najjar, R.~R\"odl, A.~Sch\"afer, R.~Schiel,
W.~S\"oldner, \speaker{A.~Sternbeck} ~and P.~Wein\\
Institut f\"ur Theoretische Physik, Universit\"at Regensburg, 93040 Regensburg,
Germany\\
E-mail: \email{andre.sternbeck@ur.de}}

\abstract{We present new $N_f=2$ data for the nucleon generalized form
factors, varying volume, lattice spacing and pion mass, down to 150 MeV.
We also give an update of our direct calculation of the
nucleon sigma term for a range of pion mass values including the lightest one.}

\FullConference{31st International Symposium on Lattice Field Theory - LATTICE 2013\\
		July 29 - August 3, 2013\\
		Mainz, Germany}

\begin{document}

\section*{Introduction}

Lattice QCD calculations are an essential tool to study the inner structure of
nucleons. Moments of parton distribution functions and form factors or, more
recently, generalized form factors have always been popular targets in this
regard. These observables provide important information on the distribution of
momentum, spin and charge within a nucleon and are accessible on the lattice via
expectation values of local operators, $O^{\mu_1\mu_2,\ldots}$. For example,
\begin{equation}
  \label{eq:Tmunu}
  \langle N(P')|O^{\mu\nu}_V|N(P)\rangle = \overline{U}(P')\bigg\{\gamma^{\{\mu}
\overline P^{\nu\}}
    A_{20}(t) - \frac{
i\sigma^{\rho\{\mu}\Delta_{\rho}\overline{P}^{\nu\}}}{2m_N} B_{20}(t)
    +\frac{\Delta^{\{\mu}\Delta^{\nu\}}}{m_N} C_{2}(t) \bigg\}U(P) 
\end{equation}
gives access to the twist-2 generalized form factors $A_{20}$, $B_{20}$
and $C_2$ and also to the second (Mellin) moment of the nucleon's parton
distribution $\langle x\rangle=A_{20}(0)$ (see, e.g., \cite{Hagler:2009ni} for a
comprehensive list).

Besides structure functions, sigma terms, $\sigma_q = m_q\langle N\vert
\bar{q}q \vert N\rangle$ for the quark $q=\{u,d,s\}$, are also of interest. They
parametrize the (small) contribution of quarks to the nucleon mass and are
needed, e.g., for precision measurements of SM parameters or
dark-matter searches. Experimentally, sigma terms are only indirectly accessible
but lattice QCD provides a way to calculate them. In particular, the sigma term
of the strange quark, $\sigma_s$, or the so-called pion-nucleon sigma term
$\sigma_{\pi N} = \sigma_u + \sigma_d$ of the (approximate mass-degenerate)
light quarks have been addressed on the lattice in the past
\cite{Babich:2010at,Bali:2011ks,Dinter:2012tt}. Until recently it was however
numerically too expensive to perform these calculations in the vicinity of the
physical point or to directly evaluate $\sigma_{\pi N}$. Therefore, most lattice
values of $\sigma_{\pi N}$ are from indirect determinations, for example, from
chiral extrapolations of lattice data for the nucleon mass (via the
Feynman-Hellmann theorem) \cite{Young:2009zb,*Horsley:2011wr,*Durr:2011mp,
*Shanahan:2012wh,*Chowdhury:2012wa,*Alvarez-Ruso:2013fza,Bali:2012qs}.
Often these extrapolations are performed up to relatively large pion masses
$m_\pi$, which limits the precision of the final (physical) value.

Lattice determinations of structure functions suffer similar limitations, in
addition to the problem of excited-state contributions which become more severe
the closer one gets to the physical point. With the advance of new lattice
techniques and ever more powerful computers the situation has, however,
improved. Now lattice QCD calculations start to approach the physical point and
excited-state contaminations can be removed more efficiently
\cite{Bali:2013nla}. Here we will present new data for the generalized form
factors and $\sigma_{\pi N}$, including an estimate almost at the physical
point. 

\section*{Lattice setup}

Our lattice calculations are performed on configurations generated by the
Regensburg group and QCDSF of $N_f=2$ nonperturbatively-improved Wilson
fermions and the standard Wilson gauge action with $\beta=5.29$ and 5.40. The
lattice spacings are 0.07 and 0.06\,fm, respectively and our pion mass values
range from 491\,MeV down to 150\,MeV. Lattice sizes are chosen accordingly:
$Lm_\pi\ge3.4$ or higher (see \Tab{tab:simpara} for details). For the
translation of our lattice results into physical units we assume
$r_0=0.5\,\text{fm}$ \cite{Bali:2012qs} and use nonperturbative renormalization
constants for the conversion of the form factors to the
$\mathrm{\overline{MS}}$ scheme (renormalization scale $\mu=2\,\text{GeV}$)
\cite{Gockeler:2010yr}. 

Our calculations feature an optimized source and sink smearing to reduce
the impact of excited-state contaminations. These sources and sinks are created
using typically 300--400 steps of Wuppertal smearing and APE-smeared spatial
links. Two or three sources are used per configuration, depending on the
ensemble parameters (see \Tab{tab:simpara}). The first source is placed
at a random site, the remaining one (or two) such that the distance between
them is maximized. 

The generalized form factors (GFFs) are extracted from fits to ratios
\begin{align} 
 \label{eq:ratio}
  R(t_{sink},\tau,p',p) =
 \frac{C^{\mathcal{O}}_{3pt}(t_{sink},\tau,\vec{p}',\vec{p})}{C_{2pt}(t_{sink},
\vec{p}')}
\sqrt
 {\frac{C_{2pt}(t_{sink}-\tau,\vec{p})C_{2pt}(\tau,\vec{p}')C_{2pt}(t_{sink},
\vec{p}')}{C_{2pt}(t_{sink}-\tau,\vec{p}')C_{2pt}(\tau,\vec{p})C_{2pt}(t_{sink},
\vec{p})}}
\end{align}
of nucleon two and three-point functions, $C_{2pt}(t,\vec {p})$ and
$C_{3pt}^\mathcal{O}(t_{sink},\tau,\vec {p}',\vec {p})$. For the
calculation of the latter we use the standard sequential-source
technique with a current inserted at \hbox{$0\ll\tau\ll t_{sink}$} (see
\cite{Bali:2013gxx} for our calculation using stochastic
noise). $t_{sink}$ denotes the source-sink separation. We find
that $t_{sink}=15$ and $t_{sink}=17$ (in lattice units) are sufficient for
our calculations at $\beta=5.29$ and $\beta=5.40$, respectively. For tests
using different $t_{sink}$ values see \cite{Bali:2013nla}. We restrict
ourselves to the isovector case, but disconnected contributions are in
progress.

For the scalar forward matrix element we have already preliminary data,
including disconnected contributions. We extend our former analysis of
$\sigma_{\pi N}$ \cite{Bali:2011ks} to a range of pion mass values. The scalar
matrix elements $\langle N\vert \bar{q}q \vert N\rangle$, from
which we obtain estimates for $\sigma_{\pi N}$, are extracted from ratios
of three- and two-point functions at zero lattice momentum and with a stochastic
estimator for the disconnected contributions.

\begin{table}
\begin{center}
\begin{tabular}{ccc@{\quad}rcccc}\hline\hline
$\beta$ & $\kappa$ & lattice  & $N\times M$ & $a$~[fm] &  $m_{\pi}$~[MeV]
& $Lm_\pi$ & $t_{sink}$/a\\
\hline
$5.29$    &  0.13620 & $24^3\times 48$  & $1124\times 2$ & 0.07 &  430  & 3.7
& 15  \\
 & 0.13632&  $32^3\times 64$  & $2027\times 2$ &  & 294   & 3.4  & 15
\\
 & 0.13632&  $40^3\times 64$  & $2028\times 2$ &  & 289   & 4.2  & 15\\
 & 0.13640&  $64^3\times 64$  &  $940\times 3$ &  & 150   & 3.5  &
15\\*[1ex]

$5.40$ & 0.13640 & $32^3\times 64$ & $1170\times 2$ & 0.06 & 491 & 4.8 &
17\\
      & 0.13660 & $48^3\times 64$ & $2178\times 2$ & & 260 & 3.8 &
17\\\hline
\end{tabular}
\end{center}
\caption{Simulation parameters. $N$ is the number of configurations
and $M$ of sources per configuration.}
\label{tab:simpara}
\end{table} 

\section*{Results for Generalized Form Factors}

\begin{figure*}
 \mbox{\includegraphics[width=0.5\linewidth]{%
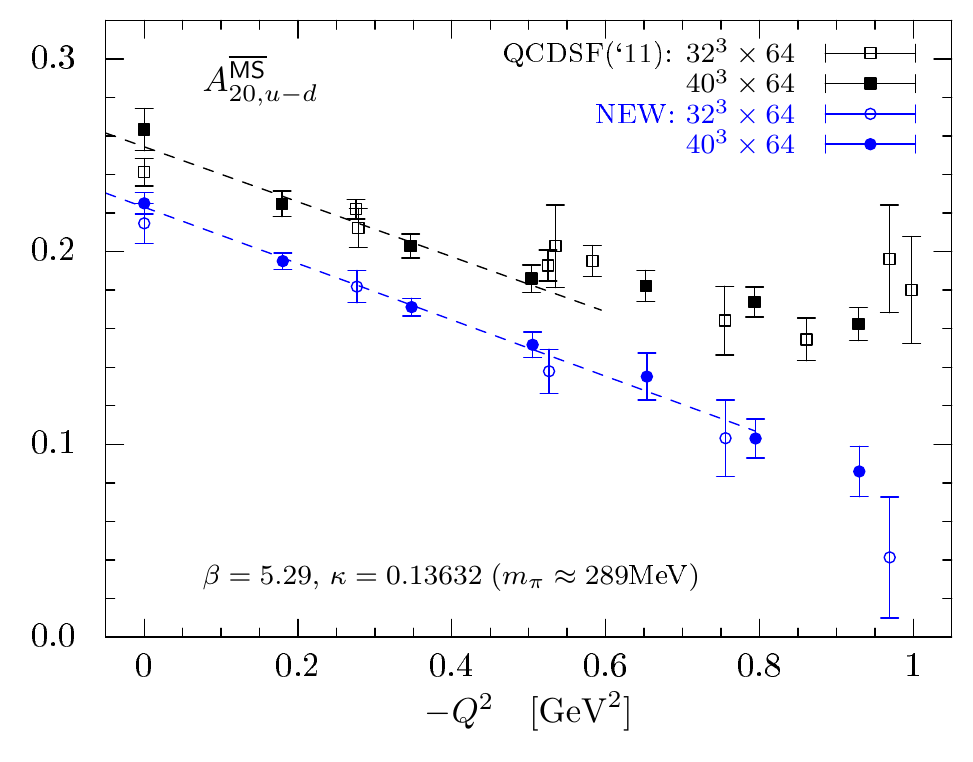}
 \includegraphics[width=0.5\linewidth]{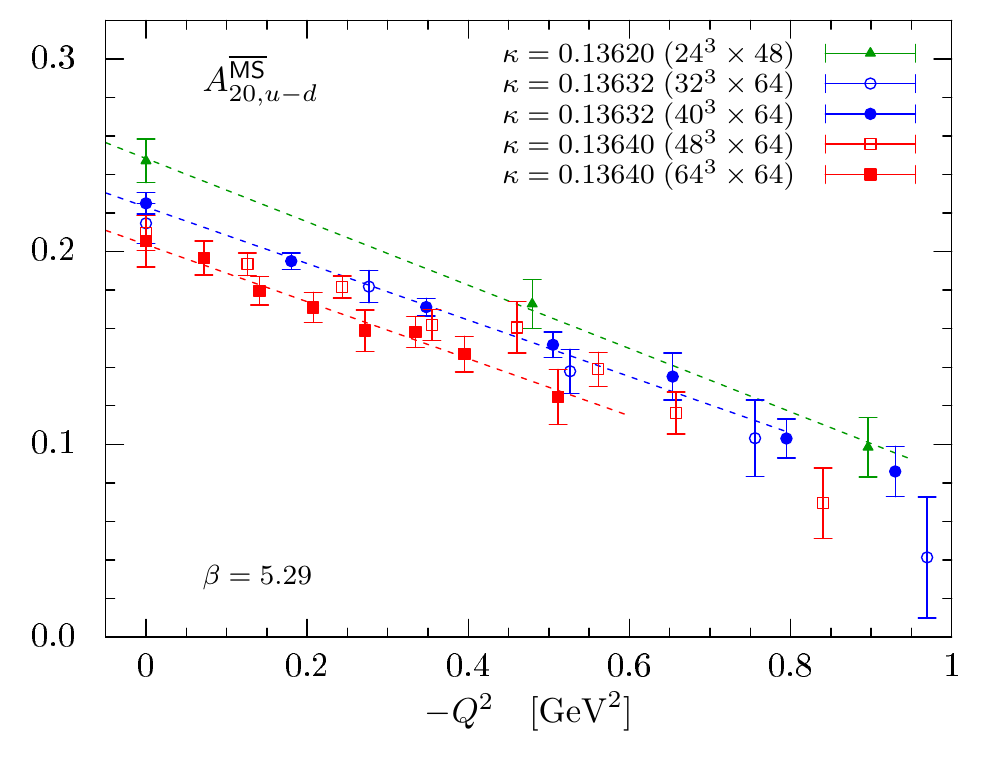}}
\caption{$A^{\mathsf{\overline{MS}}}_{20,u-d}$ versus $-Q^2$. \textbf{Left:}
comparison of two types of source smearing for fixed lattice parameters
($\beta=5.29$, $\kappa=0.13632$). Black squares are for Jacobi  smearing (old
data from \cite{Sternbeck:2012rw}), blue circles for the improved
smearing. \textbf{Right:} $Q^2$-dependence for three pion
masses: $m_\pi=151$ (red squares), 289 (blue circles) and  429\,MeV (green
diamonds). The points are all for the improved smearing and $\beta=5.29$.}
\label{fig:a20_umd_b5p29}
\bigskip
\mbox{\includegraphics[width=0.5\linewidth]{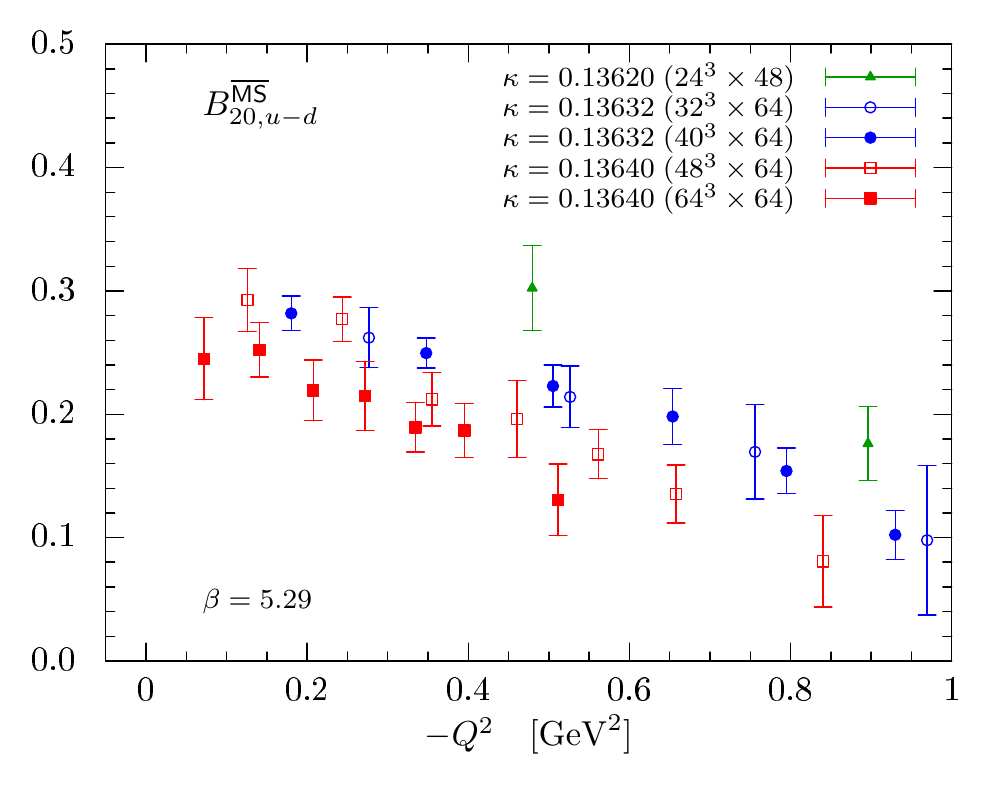}
 \includegraphics[width=0.5\linewidth]{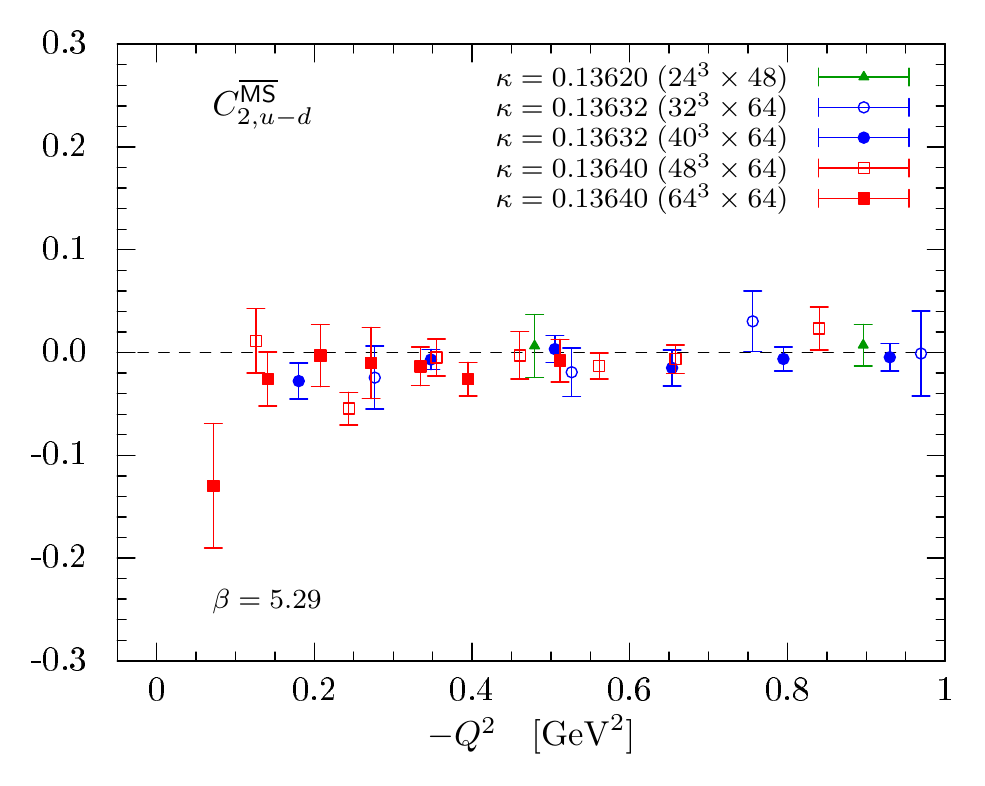}}
\mbox{\includegraphics[width=0.5\linewidth]{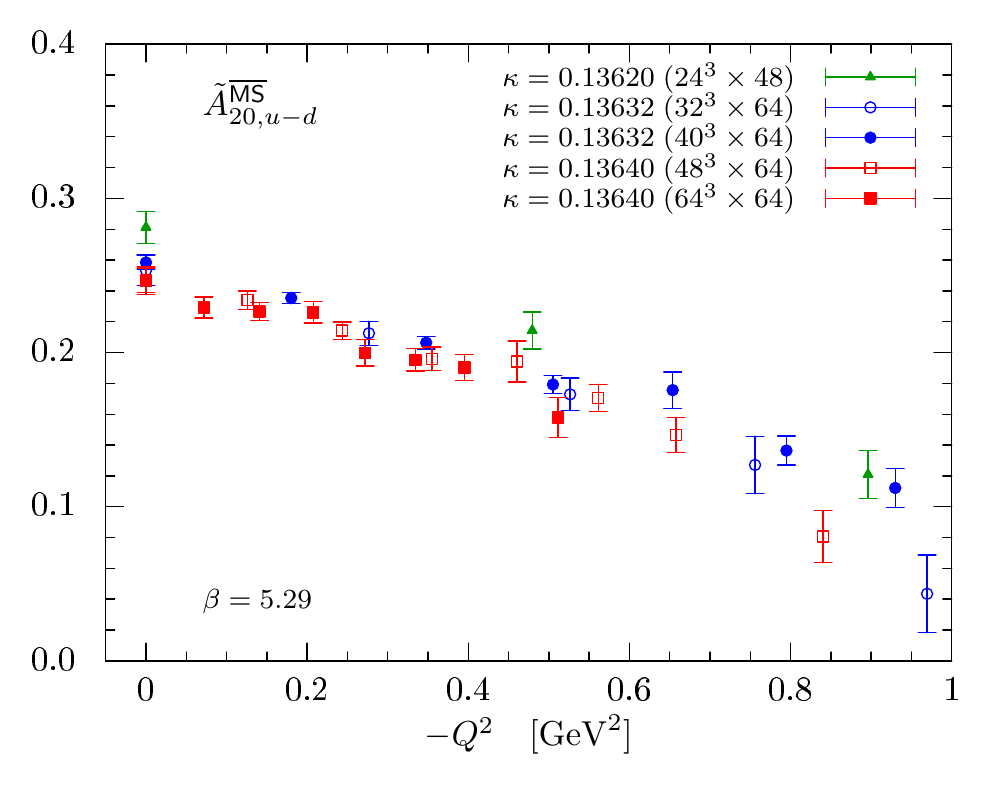}
 \includegraphics[width=0.5\linewidth]{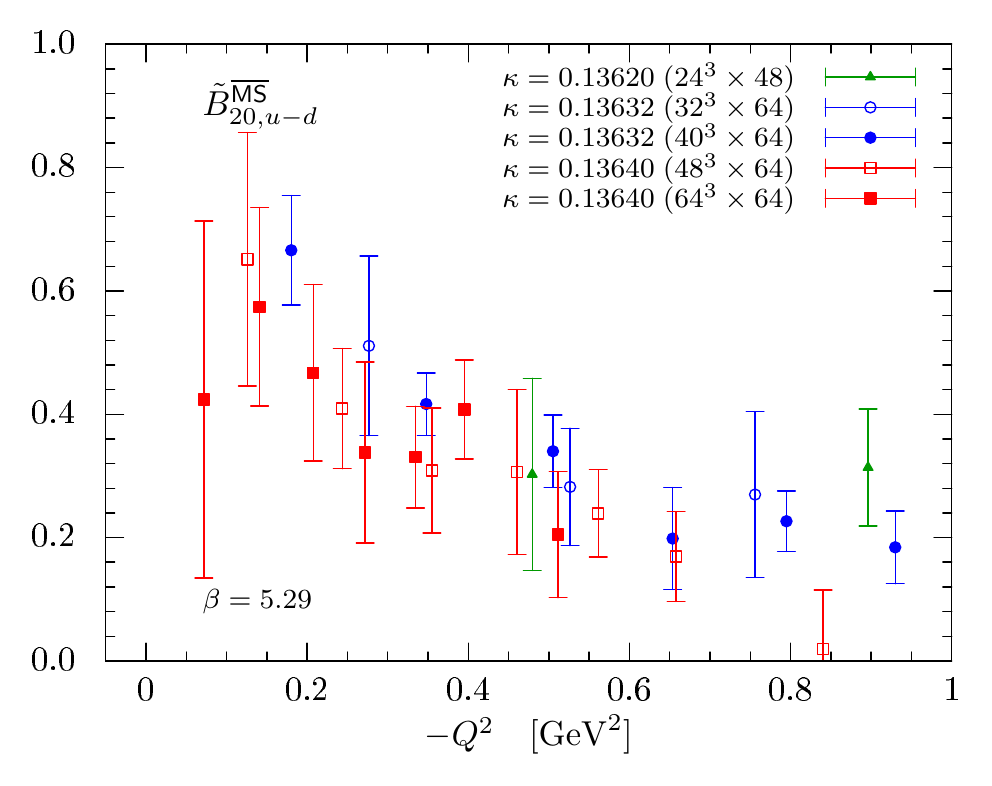}}
\caption{As \protect\Fig{fig:a20_umd_b5p29}
(right) but for
$B^{\mathsf{\overline{MS}}}_{20,u-d}$ (top left),
$C^{\mathsf{\overline{MS}}}_{2,u-d}$
(top right), $\tilde{A}^{\mathsf{\overline{MS}}}_{20,u-d}$ (bottom left)
and $\tilde{B}^{\mathsf{\overline{MS}}}_{20,u-d}$ (bottom right). Note that all
data are still preliminary.}
 \label{fig:b20_etal_umd_b5p29}
\end{figure*}

We start the discussion of results with the GFFs, namely the second (Mellin)
moments of the vector and axial-vector nucleon GPDs.\footnote{Results for the
first and third moments will be published elsewhere. Pion structure function
results from the same gauge ensembles can be found in \cite{Bali:2013gya}.} The
relevant form factors are $A_{20}$, $B_{20}$ and $C_{2}$ for the vector GPDs,
and $\tilde{A}_{20}$ and $\tilde{B}_{20}$ for the axial-vector GPDs. Preliminary
data for $A_{20}$, $B_{20}$ and $C_{2}$ were already shown in
\cite{Sternbeck:2012rw}. Here we provide an update for these (obtained
with our improved sink-source smearing) and also present results from a new
gauge ensemble at almost physical quark mass ($m_\pi\approx150\,$MeV).

A comparison of new and old points is shown in \Fig{fig:a20_umd_b5p29} (left)
for the example of $A^{\mathsf{\overline{MS}}}_{20,u-d}$. Our new points lie
systematically below the old points and this difference is almost independent of
the momentum transfer $Q^2$, and roughly of the same magnitude as we see for
$\langle x \rangle_{u-d} = A^{\mathsf{\overline{MS}}}_{20,u-d}(0)$~ (see
\cite{Bali:2013nla}). Since all data sets originate from the same gauge
ensembles ($m_\pi\approx289\,$MeV), but a different type of sink-source
smearing was used, we believe our improved smearing is the main reason for
the down-shift of our new points, as has been discussed in detail in
\cite{Bali:2013nla}.

Deviations are also seen for $B_{20,u-d}$, but here these deviations decreases
with \hbox{$\vert Q^2\vert\to 0$}. In fact, the slopes of the new and old points
differ, but towards $Q^2=0$ the points tend to the same values. For $C_{20,u-d}$
we practically see no difference between old and new points.

Looking at the $m_\pi$-dependence, we see this dependence is now more pronounced
than for the old data. For example, for $A_{20,u-d}(Q^2)$ we see that points
for $Q^2<0$ differ for different pion masses, while in \cite{Sternbeck:2012rw}
we saw this difference to decrease with $Q^2\to 0$ (cf.\ Fig.~2 of
\cite{Sternbeck:2012rw} to \Fig{fig:a20_umd_b5p29} (right)).  Also for 
$B_{20,u-d}$ and $\tilde{A}_{20}$ we observe a vertical separation of points for
different pion masses that does not disappear as $Q^2\to 0$.

When the full statistics is reached we should be able to provide precise
data for the low-$Q^2$ dependence of the GFFs. We will also perform combined
fits of these five form factors to new (full) 1-loop expressions from Baryon
Chiral perturbation theory \cite{PWein}. As mentioned, calculations including
disconnected contributions are in progress. 

\section*{Results for the Nucleon Sigma term}

\begin{figure*}
 \includegraphics[width=0.9\linewidth]{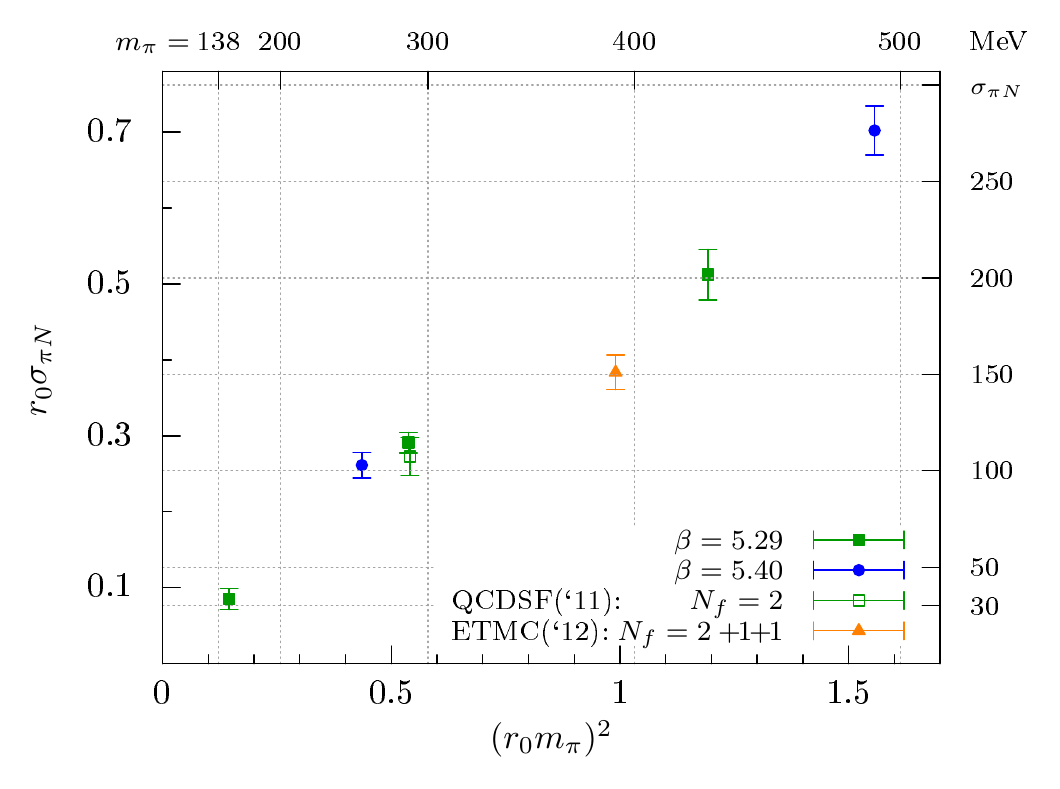}
 \caption{The $N_f=2$ pion-nucleon sigma term versus $m_\pi^2$.
  The ETMC point is from \cite{Dinter:2012tt} and the QCDSF point from
\cite{Bali:2011ks}. The right y-axis and top x-axis give the corresponding
physical numbers.}
 \label{fig:sigma_vs_mpi2}
\bigskip
 \includegraphics[width=0.8\linewidth]{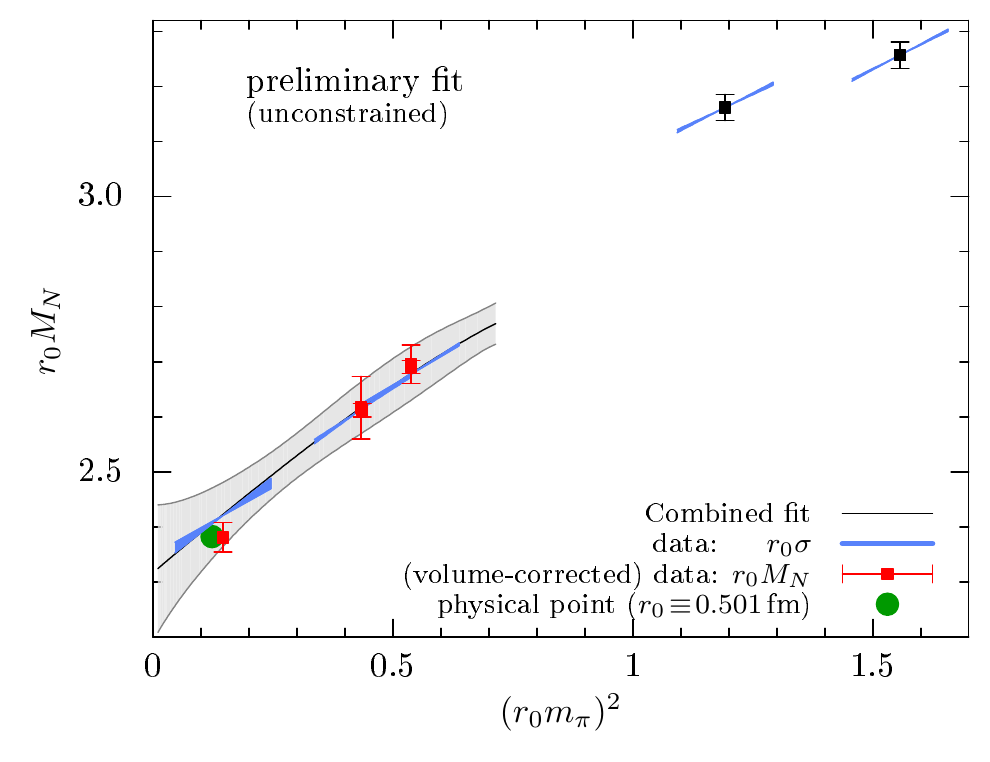}
 \caption{Combined fit to the (volume-corrected) nucleon mass data (red squares)
  and sigma term data (blue slopes), similar to what we did in
\cite{Bali:2012qs}. The physical point is marked by a green circle assuming
$r_0=0.501\,\textrm{fm}$ \cite{Bali:2012qs}. The fit was not
\emph{``constrained''} to go through the physical point.}
 \label{fig:Mn_vs_mpi2}
\end{figure*}

For the scalar matrix element we already have some preliminary results which
include contributions from disconnected diagrams. This allows us to update
our former data on $\sigma_{\pi N}$. Our new $\sigma_{\pi N}$ data are shown
in \Fig{fig:sigma_vs_mpi2} for pion masses between $150\,\textrm{MeV}$ and
$491\,\textrm{MeV}$. There we also show our previous estimate at
$m_\pi\approx290\,\textrm{MeV}$ \cite{Bali:2011ks} and the $N_f=2\!+\!1\!+\!1$
point from the ETM collaboration at $m_\pi\approx 390\,\textrm{MeV}$
\cite{Dinter:2012tt}. Both agree within errors with our new ($N_f=2$) points. 

As mentioned above, at the physical point one knows $\sigma^{phys}_{\pi
N}$ only indirectly, for example from chiral extrapolations of
lattice data for the nucleon mass. Recent values for $\sigma^{phys}_{\pi N}$
from such studies range between 32 and 52\,MeV
\cite{Young:2009zb,*Horsley:2011wr,*Durr:2011mp,*Shanahan:2012wh,%
*Chowdhury:2012wa,*Alvarez-Ruso:2013fza,Bali:2012qs}. Our new direct (but still
preliminary) data point at $m_\pi\approx 150\,\text{MeV}$ ($a\approx
0.07\,\text{fm}$) lies in the lower half of that range (see
\Fig{fig:sigma_vs_mpi2}). 

With these new points for $\sigma_{\pi N}$ we can also refine our scale
setting of $r_0$ for which in \cite{Bali:2012qs} we used nucleon mass data for
a range of pion mass values below 500\,MeV and the sigma-term at
$m_\pi\approx290\,\textrm{MeV}$ and fitted them simultaneously to the chiral
expressions for $\sigma_{\pi N}(m^2_\pi)$ and $M_N(m^2_\pi)$. Since $\sigma_{\pi
N}(m^2_\pi)$ is related to the slope of $M_N$, this particular
combination helps to reduce the uncertainties of the chiral fits and hence of
the scale setting. In \cite{Bali:2012qs} we had $\sigma_{\pi N}$ only for a
single $m_\pi$, but now we have both $\sigma_{\pi N}(m^2_\pi)$ and
$M_N(m^2_\pi)$ for all $m_\pi$.

An example for such a simultaneous fit (which also includes volume corrections)
is shown in \Fig{fig:Mn_vs_mpi2}. There, the (blue) slopes are given through the
$\sigma_{\pi N}(m^2_\pi)$ data and the (red) squares are the finite-volume
corrected data points for the nucleon mass. This fit is still preliminary and
only shown for a particular pion mass range, but it demonstrates the potential
of our approach.

Note also the physical point marked by a (green) full circle in
\Fig{fig:Mn_vs_mpi2}. For this we use our estimate for $r_0$ from
\cite{Bali:2012qs}, which is clearly supported by our new data. The fit
was not forced to go through the physical point.

\section*{Conclusions}

We have reported on our reanalysis of the nucleon generalized form factors
(GFFs) on $N_f=2$ gauge field ensembles and added a new ensemble with
$m_\pi\approx 150\,\textrm{MeV}$. Our calculations feature an optimized source
and sink smearing for which we find excited-state contaminations are much
reduced (see also \cite{Bali:2013nla}). A comparison of our new and old data
for the GFFs shows a better removal of excited state contributions that changes
the $Q^2$ dependence. In particular the $m_\pi$-dependence is much more
pronounced. 

Along with this re-calculation of the GFFs, we have also extended our
calculation of $\sigma_{\pi N}$. We have provided here new $N_f=2$ data for
$\sigma_{\pi N}(m_\pi)$ for pion masses from $m_\pi\approx 491\,\textrm{MeV}$
down to $m_\pi\approx 150\,\textrm{MeV}$, which will allow us to give an
improved estimate of $\sigma^{phys}_{\pi N}$ in the near future.

The presented results should be considered as still preliminary. A complete
analysis will follow as soon as the full statistics has been reached for all
gauge ensembles.

{\small
\bigskip
\begin{spacing}{1.1}
 Numerical calculations have been performed on the SuperMUC system at the
LRZ/Germany and the FERMI BG/Q machine at CINECA/Italy. We acknowledge PRACE
(project 2011050791) for awarding us access to the FERMI BG/Q machine. 
We have made use of the Chroma software suite \cite{Edwards:2004sx} adapted for
our needs. For the generation of gauge field configurations we used QPACE and 
the BQCD software \cite{Nakamura:2010qh} including an improved
inverter \cite{Nobile:2010zz}. This work has been supported in part by the DFG
(SFB/TR 55, Hadron Physics from Lattice QCD) and the EU under grant 238353 (ITN
STRONGnet). A.St acknowledges support by the European Reintegration Grant
(FP7-PEOPLE-2009-RG, No.256594).
\end{spacing}
}

\begin{thebibliography}{10}%
\makeatletter
\providecommand \@ifxundefined [1]{%
 \ifx #1\undefined \expandafter \@firstoftwo
 \else \expandafter \@secondoftwo
\fi
}%
\providecommand \@ifnum [1]{%
 \ifnum #1\expandafter \@firstoftwo
 \else \expandafter \@secondoftwo
\fi
}%
\providecommand \enquote [1]{``#1''}%
\providecommand \bibnamefont  [1]{#1}%
\providecommand \bibfnamefont [1]{#1}%
\providecommand \citenamefont [1]{#1}%
\providecommand\href[0]{\@sanitize\@href}%
\providecommand\@href[1]{\endgroup\@@startlink{#1}\endgroup\@@href}%
\providecommand\@@href[1]{#1\@@endlink}%
\providecommand \@sanitize [0]{\begingroup\catcode`\&12\catcode`\#12\relax}%
\@ifxundefined \pdfoutput {\@firstoftwo}{%
 \@ifnum{\z@=\pdfoutput}{\@firstoftwo}{\@secondoftwo}%
}{%
 \providecommand\@@startlink[1]{\leavevmode}%
 \providecommand\@@endlink[0]{}%
}{%
 \providecommand\@@startlink[1]{%
  \leavevmode
  \pdfstartlink
   attr{/Border[0 0 1 ]/H/I/C[0 1 1]}%
   user{/Subtype/Link/A<</Type/Action/S/URI/URI(#1)>>}%
  \relax
 }%
 \providecommand\@@endlink[0]{\pdfendlink}%
}%
\providecommand \url  [0]{\begingroup\@sanitize \@url }%
\providecommand \@url [1]{\endgroup\@href {#1}{\urlprefix}}%
\providecommand \urlprefix [0]{URL }%
\providecommand \Eprint[0]{\href }%
\@ifxundefined \urlstyle {%
  \providecommand \doi [1]{doi:\discretionary{}{}{}#1}%
}{%
  \providecommand \doi [0]{doi:\discretionary{}{}{}\begingroup
  \urlstyle{rm}\Url }%
}%
\providecommand \doibase [0]{http://dx.doi.org/}%
\providecommand \Doi[1]{\href{\doibase#1}}%
\providecommand \bibAnnote [3]{%
  \BibitemShut{#1}%
  \begin{quotation}\noindent
    \textsc{Key:}\ #2\\\textsc{Annotation:}\ #3%
  \end{quotation}%
}%
\providecommand \bibAnnoteFile [2]{%
  \IfFileExists{#2}{\bibAnnote {#1} {#2} {\input{#2}}}{}%
}%
\providecommand \typeout [0]{\immediate \write \m@ne }%
\providecommand \selectlanguage [0]{\@gobble}%
\providecommand \bibinfo [0]{\@secondoftwo}%
\providecommand \bibfield [0]{\@secondoftwo}%
\providecommand \translation [1]{[#1]}%
\providecommand \BibitemOpen[0]{}%
\providecommand \bibitemStop [0]{}%
\providecommand \bibitemNoStop [0]{.\EOS\space}%
\providecommand \EOS [0]{\spacefactor3000\relax}%
\providecommand \BibitemShut [1]{\csname bibitem#1\endcsname}%
\bibitem{Hagler:2009ni}%
  \BibitemOpen
  \bibfield{author}{%
  \bibinfo {author} {\bibfnamefont{P.}~\bibnamefont{H{\"a}gler}},\ }%
  \bibfield{journal}{%
  \Doi{10.1016/j.physrep.2009.12.008}{\bibinfo {journal} {Phys.Rept.}}\ }%
  \textbf{\bibinfo {volume} {490}},\ \bibinfo {pages} {49} (\bibinfo {year}
  {2010}),\ \Eprint{http://arxiv.org/abs/0912.5483}{arXiv:0912.5483 [hep-lat]}%
  \bibAnnoteFile{NoStop}{Hagler:2009ni}%
\bibitem{Babich:2010at}%
  \BibitemOpen
  \bibfield{author}{%
  \bibinfo {author} {\bibfnamefont{R.}~\bibnamefont{Babich}} \emph{et~al.},\ }%
  \bibfield{journal}{%
  \bibinfo {journal} {Phys.Rev.}\ }%
  \textbf{\bibinfo {volume} {D85}},\ \bibinfo {pages} {054510} (\bibinfo {year}
  {2012}),\ \Eprint{http://arxiv.org/abs/1012.0562}{arXiv:1012.0562 [hep-lat]}%
  \bibAnnoteFile{NoStop}{Babich:2010at}%
\bibitem{Bali:2011ks}%
  \BibitemOpen
  \bibfield{author}{%
  \bibinfo {author} {\bibfnamefont{G.~S.}\ \bibnamefont{Bali}} \emph{et~al.},\
  }%
  \bibfield{journal}{%
  \Doi{10.1103/PhysRevD.85.054502}{\bibinfo {journal} {Phys.Rev.}}\ }%
  \textbf{\bibinfo {volume} {D85}},\ \bibinfo {pages} {054502} (\bibinfo {year}
  {2012}),\ \Eprint{http://arxiv.org/abs/1111.1600}{arXiv:1111.1600 [hep-lat]}%
  \bibAnnoteFile{NoStop}{Bali:2011ks}%
\bibitem{Dinter:2012tt}%
  \BibitemOpen
  \bibfield{author}{%
  \bibinfo {author} {\bibfnamefont{S.}~\bibnamefont{Dinter}} \emph{et~al.}
  (\bibinfo {collaboration} {ETM Collaboration}),\ }%
  \bibfield{journal}{%
  \Doi{10.1007/JHEP08(2012)037}{\bibinfo {journal} {JHEP}}\ }%
  \textbf{\bibinfo {volume} {1208}},\ \bibinfo {pages} {037} (\bibinfo {year}
  {2012}),\ \Eprint{http://arxiv.org/abs/1202.1480}{arXiv:1202.1480 [hep-lat]}%
  \bibAnnoteFile{NoStop}{Dinter:2012tt}%
\bibitem{Young:2009zb}%
  \BibitemOpen
  \bibfield{author}{%
  \bibinfo {author} {\bibfnamefont{R.}~\bibnamefont{Young}}\ and\ \bibinfo
  {author} {\bibfnamefont{A.}~\bibnamefont{Thomas}},\ }%
  \bibfield{journal}{%
  \Doi{10.1103/PhysRevD.81.014503}{\bibinfo {journal} {Phys.Rev.}}\ }%
  \textbf{\bibinfo {volume} {D81}},\ \bibinfo {pages} {014503} (\bibinfo {year}
  {2010}),\ \Eprint{http://arxiv.org/abs/0901.3310}{arXiv:0901.3310 [hep-lat]}%
  \bibAnnoteFile{NoStop}{Young:2009zb}%
\bibitem{Horsley:2011wr}%
  \BibitemOpen
  \bibfield{author}{%
  \bibinfo {author} {\bibfnamefont{R.}~\bibnamefont{Horsley}} \emph{et~al.}
  (\bibinfo {collaboration} {QCDSF-UKQCD Collaborations}),\ }%
  \bibfield{journal}{%
  \Doi{10.1103/PhysRevD.85.034506}{\bibinfo {journal} {Phys.Rev.}}\ }%
  \textbf{\bibinfo {volume} {D85}},\ \bibinfo {pages} {034506} (\bibinfo {year}
  {2012}),\ \Eprint{http://arxiv.org/abs/1110.4971}{arXiv:1110.4971 [hep-lat]}%
  \bibAnnoteFile{NoStop}{Horsley:2011wr}%
\bibitem{Durr:2011mp}%
  \BibitemOpen
  \bibfield{author}{%
  \bibinfo {author} {\bibfnamefont{S.}~\bibnamefont{D{\"u}rr}} \emph{et~al.},\
  }%
  \bibfield{journal}{%
  \Doi{10.1103/PhysRevD.85.014509}{\bibinfo {journal} {Phys.Rev.}}\ }%
  \textbf{\bibinfo {volume} {D85}},\ \bibinfo {pages} {014509} (\bibinfo {year}
  {2012}),\ \Eprint{http://arxiv.org/abs/1109.4265}{arXiv:1109.4265 [hep-lat]}%
  \bibAnnoteFile{NoStop}{Durr:2011mp}%
\bibitem{Shanahan:2012wh}%
  \BibitemOpen
  \bibfield{author}{%
  \bibinfo {author} {\bibfnamefont{P.}~\bibnamefont{Shanahan}}, \bibinfo
  {author} {\bibfnamefont{A.}~\bibnamefont{Thomas}},\ and\ \bibinfo {author}
  {\bibfnamefont{R.}~\bibnamefont{Young}},\ }%
  \bibfield{journal}{%
  \bibinfo {journal} {Phys.Rev.}\ }%
  \textbf{\bibinfo {volume} {D87}},\ \bibinfo {pages} {074503} (\bibinfo {year}
  {2013}),\ \Eprint{http://arxiv.org/abs/1205.5365}{arXiv:1205.5365 [nucl-th]}%
  \bibAnnoteFile{NoStop}{Shanahan:2012wh}%
\bibitem{Chowdhury:2012wa}%
  \BibitemOpen
  \bibfield{author}{%
  \bibinfo {author} {\bibfnamefont{A.}~\bibnamefont{Chowdhury}} \emph{et~al.},\
  }%
  \bibfield{journal}{%
  \Doi{10.1016/j.nuclphysb.2013.02.013}{\bibinfo {journal} {Nucl.Phys.}}\ }%
  \textbf{\bibinfo {volume} {B871}},\ \bibinfo {pages} {82} (\bibinfo {year}
  {2013}),\ \Eprint{http://arxiv.org/abs/1212.0717}{arXiv:1212.0717 [hep-lat]}%
  \bibAnnoteFile{NoStop}{Chowdhury:2012wa}%
\bibitem{Alvarez-Ruso:2013fza}%
  \BibitemOpen
  \bibfield{author}{%
  \bibinfo {author} {\bibfnamefont{L.}~\bibnamefont{Alvarez-Ruso}}, \bibinfo
  {author} {\bibfnamefont{T.}~\bibnamefont{Ledwig}}, \bibinfo {author}
  {\bibfnamefont{J.}~\bibnamefont{Martin~Camalich}},\ and\ \bibinfo {author}
  {\bibfnamefont{M.}~\bibnamefont{Vicente-Vacas}},\ }%
  \bibfield{journal}{%
  \Doi{10.1103/PhysRevD.88.054507}{\bibinfo {journal} {Phys.Rev.}}\ }%
  \textbf{\bibinfo {volume} {D88}},\ \bibinfo {pages} {054507} (\bibinfo {year}
  {2013}),\ \Eprint{http://arxiv.org/abs/1304.0483}{arXiv:1304.0483 [hep-ph]}%
  \bibAnnoteFile{NoStop}{Alvarez-Ruso:2013fza}%
\bibitem{Bali:2012qs}%
  \BibitemOpen
  \bibfield{author}{%
  \bibinfo {author} {\bibfnamefont{G.}~\bibnamefont{Bali}} \emph{et~al.},\ }%
  \bibfield{journal}{%
  \Doi{10.1016/j.nuclphysb.2012.08.009}{\bibinfo {journal} {Nucl.Phys.}}\ }%
  \textbf{\bibinfo {volume} {B866}},\ \bibinfo {pages} {1} (\bibinfo {year}
  {2013}),\ \Eprint{http://arxiv.org/abs/1206.7034}{arXiv:1206.7034 [hep-lat]}%
  \bibAnnoteFile{NoStop}{Bali:2012qs}%
\bibitem{Bali:2013nla}%
  \BibitemOpen
  \bibfield{author}{%
  \bibinfo {author} {\bibfnamefont{G.}~\bibnamefont{Bali}} \emph{et~al.},\ }%
  \bibfield{journal}{%
  \bibinfo {journal} {PoS}\ }%
  \textbf{\bibinfo {volume} {Lattice 2013}},\ \bibinfo {pages} {290} (\bibinfo
  {year} {2013}),\ \Eprint{http://arxiv.org/abs/1311.7041}{arXiv:1311.7041
  [hep-lat]}%
  \bibAnnoteFile{NoStop}{Bali:2013nla}%
\bibitem{Gockeler:2010yr}%
  \BibitemOpen
  \bibfield{author}{%
  \bibinfo {author} {\bibfnamefont{M.}~\bibnamefont{G{\"o}ckeler}}
  \emph{et~al.},\ }%
  \bibfield{journal}{%
  \Doi{10.1103/PhysRevD.82.114511}{\bibinfo {journal} {Phys.Rev.}}\ }%
  \textbf{\bibinfo {volume} {D82}},\ \bibinfo {pages} {114511} (\bibinfo {year}
  {2010}),\ \Eprint{http://arxiv.org/abs/1003.5756}{arXiv:1003.5756 [hep-lat]}%
  \bibAnnoteFile{NoStop}{Gockeler:2010yr}%
\bibitem{Bali:2013gxx}%
  \BibitemOpen
  \bibfield{author}{%
  \bibinfo {author} {\bibfnamefont{G.~S.}\ \bibnamefont{Bali}} \emph{et~al.}}%
   (\bibinfo {year} {2013}),\
  \Eprint{http://arxiv.org/abs/1311.1718}{arXiv:1311.1718 [hep-lat]}%
  \bibAnnoteFile{NoStop}{Bali:2013gxx}%
\bibitem{Sternbeck:2012rw}%
  \BibitemOpen
  \bibfield{author}{%
  \bibinfo {author} {\bibfnamefont{A.}~\bibnamefont{Sternbeck}} \emph{et~al.},\
  }%
  \bibfield{journal}{%
  \bibinfo {journal} {PoS}\ }%
  \textbf{\bibinfo {volume} {LATTICE2011}},\ \bibinfo {pages} {177} (\bibinfo
  {year} {2011}),\ \Eprint{http://arxiv.org/abs/1203.6579}{arXiv:1203.6579
  [hep-lat]}%
  \bibAnnoteFile{NoStop}{Sternbeck:2012rw}%
\bibitem{Bali:2013gya}%
  \BibitemOpen
  \bibfield{author}{%
  \bibinfo {author} {\bibfnamefont{G.}~\bibnamefont{Bali}} \emph{et~al.},\ }%
  \bibfield{journal}{%
  \bibinfo {journal} {PoS}\ }%
  \textbf{\bibinfo {volume} {LATTICE 2013}},\ \bibinfo {pages} {447} (\bibinfo
  {year} {2013}),\ \Eprint{http://arxiv.org/abs/1311.7639}{arXiv:1311.7639
  [hep-lat]}%
  \bibAnnoteFile{NoStop}{Bali:2013gya}%
\bibitem{PWein}%
  \BibitemOpen
  \bibfield{author}{%
  \bibinfo {author} {\bibfnamefont{P.}~\bibnamefont{Wein}} \emph{et~al.},\ }%
  \bibinfo {note} {work in progress}%
  \bibAnnoteFile{NoStop}{PWein}%
\bibitem{Edwards:2004sx}%
  \BibitemOpen
  \bibfield{author}{%
  \bibinfo {author} {\bibfnamefont{R.~G.}\ \bibnamefont{Edwards}}\ and\
  \bibinfo {author} {\bibfnamefont{B.}~\bibnamefont{Joo}} (\bibinfo
  {collaboration} {SciDAC + LHPC + UKQCD}),\ }%
  \bibfield{journal}{%
  \Doi{10.1016/j.nuclphysbps.2004.11.254}{\bibinfo {journal}
  {Nucl.Phys.Proc.Suppl.}}\ }%
  \textbf{\bibinfo {volume} {140}},\ \bibinfo {pages} {832} (\bibinfo {year}
  {2005}),\ \Eprint{http://arxiv.org/abs/hep-lat/0409003}{arXiv:hep-lat/0409003
  [hep-lat]}%
  \bibAnnoteFile{NoStop}{Edwards:2004sx}%
\bibitem{Nakamura:2010qh}%
  \BibitemOpen
  \bibfield{author}{%
  \bibinfo {author} {\bibfnamefont{Y.}~\bibnamefont{Nakamura}}\ and\ \bibinfo
  {author} {\bibfnamefont{H.}~\bibnamefont{St{\"u}ben}},\ }%
  \bibfield{journal}{%
  \bibinfo {journal} {PoS}\ }%
  \textbf{\bibinfo {volume} {LATTICE2010}},\ \bibinfo {pages} {040} (\bibinfo
  {year} {2010}),\ \Eprint{http://arxiv.org/abs/1011.0199}{arXiv:1011.0199
  [hep-lat]}%
  \bibAnnoteFile{NoStop}{Nakamura:2010qh}%
\bibitem{Nobile:2010zz}%
  \BibitemOpen
  \bibfield{author}{%
  \bibinfo {author} {\bibfnamefont{A.}~\bibnamefont{Nobile}},\ }%
  \bibfield{journal}{%
  \bibinfo {journal} {PoS}\ }%
  \textbf{\bibinfo {volume} {LATTICE2010}},\ \bibinfo {pages} {034} (\bibinfo
  {year} {2010}),\ \Eprint{http://arxiv.org/abs/1109.4279}{arXiv:1109.4279
  [hep-lat]}%
  \bibAnnoteFile{NoStop}{Nobile:2010zz}%
\end{thebibliography}
%

\end{document}